\documentclass[11pt]{article}
\usepackage{jheppub,relsize}
\pdfoutput=1
\usepackage{slashed,ulem}
\usepackage{graphicx}
\usepackage{subfigure}
\usepackage{dcolumn}
\usepackage{bm}
\usepackage{float}
\usepackage{amsfonts}
\usepackage{xcolor}

\newcommand{\be}{\begin{equation}} 
\newcommand{\ee}{\end{equation}}  
\newcommand{\bea}{\begin{eqnarray}}  
\newcommand{\eea}{\end{eqnarray}}

\title{Gravitational Waves from a Pati-Salam Phase Transition}
\author[a]{Djuna Croon,}
\author[b,c]{Tom\'as E. Gonzalo,}
\author[a]{Graham White}

\affiliation[a]{TRIUMF Theory Group, 4004 Wesbrook Mall, Vancouver, B.C. V6T2A3, Canada}
\affiliation[b]{Department of Physics, University of Oslo, N-0316 Oslo, Norway}
\affiliation[c]{ARC Centre of Excellence for Particle Physics at the Tera-scale,
 School of Physics and Astronomy, Monash University, Melbourne, Victoria 3800, Australia }
\emailAdd{dcroon@triumf.ca}
\emailAdd{tomas.gonzalo@monash.edu}
\emailAdd{gwhite@triumf.ca}
\date{\today}
\abstract{
We analyse the gravitational wave and low energy signatures of a Pati-Salam phase transition. For a Pati-Salam scale of $M_{PS} \sim 10^5$ GeV, we find 
a stochastic power spectrum within reach of the next generation of ground-based interferometer experiments such as the Einstein Telescope, in parts of the parameter space. We study the lifetime of the proton in this model, as well as complementarity with low energy constraints including electroweak precision data, neutrino mass measurements, lepton flavour violation, and collider constraints.}

\preprint{CoEPP-MN-18-11}

\begin{document}
\maketitle
\section{Introduction}

Grand Unified Theories (GUTs)~\cite{Georgi:1974sy, Pati:1974yy, Fritzsch:1974nn, Georgi:1974my, Gursey:1975ki} are well motivated extensions of the Standard Model (SM), explaining the coincidental cancellation of SM gauge anomalies, and stabilizing the vacuum at high energy~\cite{Degrassi:2012ry}. 
GUT theories predict gauge coupling unification at a high scale $M_{GUT}\sim 10^{16}$ GeV.
Supersymmetric GUTs ~\cite{Nanopoulos:1982wk, Antoniadis:1987dx, Antoniadis:1988cm, Aulakh:1997fq, Aulakh:1982sw, Deppisch:2014aga} do so automatically~\cite{Ellis:1990wk}, but also typically require a light supersymmetric spectrum which has so far eluded experimental verification. By contrast, non-supersymmetric GUTs 
achieve gauge coupling unification through intermediate mass scales and fields that guide the gauge couplings towards unification,~\cite{Lindner:1996tf,Aulakh:2002zr,Dorsner:2005fq,Siringo:2012bc,Deppisch:2014zta,Deppisch:2014qpa,Deppisch:2017xhv}.

Gauge coupling unification in non-supersymmetric theories can therefore occur through multiple steps. One of the most well known examples of an intermediate scale model is the Pati-Salam (PS) model \cite{Pati:1974yy,Pati:2017ysg}. As opposed to fully unified models such as $SU(5)$ or $SO(10)$, PS models can survive at relatively low energies, because they do not induce rapid proton decay~\cite{Mohapatra:1980qe}. Primordial monopoles created at such low scales have been shown to be inflated away in models with a low scale strong phase transition~\cite{Senoguz:2015lba,Okada:2013vxa}. A low PS scale is phenomenologically attractive, as it implies that unification has low energy consequences which can be constrained by precision data and collider experiments. Furthermore, PS symmetry breaking can in principle lead to a first order phase transition \cite{Addazi:2018nzm}, so long as the post-inflation reheating temperature is larger than the breaking scale, which is favoured by the parameters of the theory. 

Inhomogeneous cosmic phase transitions are associated with stochastic gravitational wave (GW) spectra. Such gravitational radiation is an important mechanism through which energy is dissipated when bubbles of the new vaccuum collide. 
The associated gravitational power spectrum is therefore a function of the set of parameters which govern the thermal evolution of the phase transition: the latent heat normalized to the radiation density, $\alpha$, the speed of the transition $\beta /H$, the transition temperature $T_n$ and the velocity of the bubble wall upon collision $v_w$.
In this work, we compute the thermal parameters of a PS phase transition. We motivate an effective model with three free parameters, which describes the broken direction in the scalar potential and its most important thermal contributions. 
We study this parameter space, and show that the PS transition may lead to a stochastic spectrum which is observable in the next generation of ground-based interferometer experiments, such as the Einstein Telescope \cite{Punturo:2010zz,Christensen:2018iqi,Akutsu:2018axf,Beniwal:2018hyi,Ellis:2018mja}, and the Cosmic Explorer \cite{Evans:2016mbw}. 

Ground-based interferometer experiments are sensitive to GW spectra with relatively low PS transition scales $M_{PS}\sim \mathcal{O}(10^5)$ GeV. As such, there are several low-energy experimental directions which may probe the PS-GW parameter space.
Firstly, collider searches for right-handed neutrinos and gauge bosons become significant for low values of the scale $M_{PS}$. Moreover, further low-energy constraints may come from the neutrino sector, as the SM neutrino masses are determined by $v_L$ and $v_R$ and they have strong contributions to lepton flavour violating processes. Lastly, a GW result may inform future experimental efforts in determining the lifetime of the proton, which we have currently used to set a lower bound on $M_{PS}$.

Complementarity studies of GW from cosmic phase transitions \cite{Mazumdar:2018dfl,Weir:2017wfa,Caprini:2018mtu,Croon:2018erz,Croon:2018new,Balazs:2016tbi} have earlier focused on electroweak scale transitions in hidden sectors \cite{Jaeckel:2016jlh,Baldes:2018emh,Schwaller:2015tja,Croon:2018erz,Breitbach:2018ddu,Madge:2018gfl}, and collider signatures \cite{Dorsch:2013wja,Huang:2017jws,Chala:2018ari} and electroweak precision tests \cite{Alves:2018oct,Figueroa:2018xtu,Fujikura:2018duw} of larger Higgs sectors. The results in this work adds the study of a phase transition within a well-motivated framework, and promises a new avenue for dialogue between gravitational wave and models of particle physics.

The structure of this paper is as follows. In Section~\ref{sec:model} we describe the model, its field content, finite temperature potential and the conditions for gauge coupling unification. Section~\ref{sec:PT} contains the study of the phase transition and the spectrum of gravitational waves it produces, highlighting the optimal scenario for visibility of the GW spectrum in the next-generation of experiments. We then run this scenario through some low energy probes in Section~\ref{sec:lowenergy}, including neutrino masses, lepton flavour violation, collider searches and proton decay. We conclude in Section~\ref{sec:discussion} with a summary of the findings and a brief discussion on selected topics for expansion.
\par

\section{A first order Pati-Salam phase transition}
\label{sec:model}

The Pati-Salam model~\cite{Pati:1974yy} is a good candidate for a first order phase transition which peaks within the frequency windows of the next generation of ground-based interferometer experiments. It can admit a low energy symmetry breaking scale, $M_{PS} < 10^7$ GeV, which implies that the stochastic spectrum peaks within the experimental reach, $f_{\rm peak} \lesssim 10^3$ Hz \cite{Dev:2016feu}. The model also has a fairly large gauge coupling constant, $g_4 \gtrsim 0.8$, which increases the strength of the phase transition. Moreover, the rank of the broken group, $SU(4)$, is larger than e.g.~the electroweak phase transition, such that more latent heat is released \cite{Croon:2018erz}. 

The symmetry group of the PS model is $\mathcal{G}_{PS} = SU(4)_c \times SU(2)_L \times SU(2)_R$. This unifies the quarks and leptons of a given chirality for each generation into merely two representations of the group. The matter content in this model is therefore embedded in the representations
\begin{equation}
\{\mathbf{4},\mathbf{2},\mathbf{1}\} \leftrightarrow \left(
\begin{array}{cccc}
u_1 & u_2 & u_3 & \nu \\
d_1 & d_2 & d_3 & e
\end{array}
\right),\quad
\{\mathbf{\overline{4}},\mathbf{1},\mathbf{2^*}\} \leftrightarrow \left(
\begin{array}{cccc}
d^c_1 & d^c_2 & d^c_3 & e^c \\
-u^c_1 & -u^c_2 & -u^c_3 & -\nu^c 
\end{array}
\right).
\label{psfermions2}
\end{equation}

This gauge group and matter content are manifestly left-right symmetric, invariant under exchanges of $SU(2)_L \leftrightarrow SU(2)_R$. Manifest left-right symmetry (also known as $D$-parity) forces $g_L = g_R$, which makes gauge coupling unification an impossible task. We will thus explicitly break this symmetry by adding purely right-handed fields at the PS scale that ensure $g_L \neq g_R$ and gauge coupling unification can occur.

We therefore construct a model with a PS and a left-right (LR) symmetric group, $\mathcal{G}_{LR} = SU(3)_c \times SU(2)_L \times SU(2)_R \times U(1)_{B-L}$~\cite{Mohapatra:1974hk, Mohapatra:1974gc, Senjanovic:1975rk}, as intermediate scales from an unified UV model, e.g. $SO(10)$, with the breaking chain, 
\begin{equation}
    SO(10) \to \mathcal{G}_{PS} \to \mathcal{G}_{LR} \to \mathcal{G}_{SM},
\end{equation}
where $\mathcal{G}_{SM} = SU(3)_c \times SU(2)_L \times U(1)_{Y}$. Upon the construction of this model we aim to achieve gauge coupling unification and at the same time keep the PS breaking scale low $M_{PS} < 10^7$ GeV.

\subsection{Scalar field content}

The minimum set of scalar fields for a valid PS model needs to be sufficient to trigger spontaneous symmetry breaking (SSB) of every step in the breaking chain. This requires the following set of fields
\begin{equation}
\Phi = \{\mathbf{1},\mathbf{2},\mathbf{2}\}, \quad \Delta_R = \{\mathbf{\overline{10}},\mathbf{1},\mathbf{3}\}, \quad \Xi_1 = \{\mathbf{15},\mathbf{1},\mathbf{1}\},
\end{equation}
which trigger the symmetry breaking of $\mathcal{G}_{SM}$, $\mathcal{G}_{LR}$ and $\mathcal{G}_{PS}$ respectively. In addition we add a few more scalar fields. A left handed triplet field $\Delta_L = \{\mathbf{\overline{10}},\mathbf{3},\mathbf{1}\}$, which gives masses to the neutrinos via type II seesaw~\cite{Mohapatra:1979ia, Schechter:1980gr}. A right-handed coloured triplet, $\Omega_R = \{\mathbf{15},\mathbf{1},\mathbf{3}\}$, to explicitly break manifest LR symmetry. And two adjoint coloured fields, $\Xi_{2,3} = \{\mathbf{15},\mathbf{1},\mathbf{1}\}$ to help with gauge coupling unification.

At the scale at which the PS group is broken, when $\Xi_1$ acquires a vacuum expectation value (vev), $\langle \Xi_1 \rangle = v$, the off-diagonal gauge bosons $G$, the scalar fields $\Omega_R$, $\Xi_{2,3}$ and coloured components of $\Delta_{LR}$ get integrated out, with the following masses\footnote{See Appendix~\ref{sec:scalarpotential} for the full scalar potential of this model.}
\begin{equation}
\begin{array}{llll}
    M_G^2  &\approx \frac{1}{6} g_4^2 v^2, &\quad\quad M_{\Xi_{2,3}}^2 &\approx v^2, \\
    M_{\Omega_R}^2 &\approx \rho_1^2 v^2 - \mu_{\Omega_R}^2, & \quad\quad M_{\Delta_{L,R}^\bot}^2 &\approx v^2,
\end{array}
\label{PSmasses}
\end{equation}
where $\rho_1$ is a portal coupling between $\Omega_R$ and $\Xi_1$. The component of $\Xi_1$ that acquires the vev ($\Xi_1^v$) gets a mass $M_{\Xi_1^v} = \sqrt{2\lambda_1} v \equiv M_{PS}$, with $\lambda_1$ its quartic coupling.

After PS symmetry breaking, the remaining scalar fields decompose into representations of the LR group as
\begin{equation}
\begin{array}{llll}
    \Phi &= \{\mathbf{1},\mathbf{2},\mathbf{2}\} &\to  \phi &= \{\mathbf{1},\mathbf{2},\mathbf{2},0\}, \notag \\
    \Delta_L &= \{\mathbf{\overline{10}},\mathbf{3},\mathbf{1}\} &\to \delta_L &=  \{\mathbf{1},\mathbf{3},\mathbf{1},-2\}, \notag\\
    \notag \Delta_R &= \{\mathbf{\overline{10}},\mathbf{1},\mathbf{3}\} &\to \delta_R &= \{\mathbf{1},\mathbf{1},\mathbf{3},-2\},
\end{array}
\end{equation}

The field $\delta_R$ is now responsible for the breaking of the LR group into $\mathcal{G}_{SM}$, and the vev of $\phi$ triggers electroweak symmetry breaking. Additionally, the field $\delta_L$ acquires a vev at the same time, which has consequences for neutrino masses, as we will see later . The vevs of these fields can be expressed as
\begin{equation}
    \langle \phi \rangle = \frac{1}{\sqrt{2}} \left(\begin{matrix} v_u & 0 \\ 0 & v_d\end{matrix}\right), \quad
    \langle \delta_L \rangle = \frac{1}{\sqrt{2}} \left(\begin{matrix} 0 & 0 \\ v_L & 0\end{matrix}\right), \quad
    \langle \delta_R \rangle = \frac{1}{\sqrt{2}} \left(\begin{matrix} 0 & 0 \\ v_R & 0\end{matrix}\right).
    \label{vevs}
\end{equation}
where the SM vev is the combination $v_{SM}^2 = v_u^2 + v_d^2$. 

Lastly, after LR symmetry breaking, the gauge bosons associated with $SU(2)_R$ and $U(1)_{B-L}$, as well as the $\delta_R$ and $\delta_L$ fields get masses that look like
\begin{align}
\begin{array}{llll}
     M_{W_R}^2 &\approx \frac{1}{4}g_R^2 v_R^2, &\quad\quad\quad M_{Z_R}^2 &\approx \frac{1}{4}(g_{B-L}^2 + g_R^2)(v_{SM}^2 + 4v_R^2)\\
     M_{\delta_R}^2 &\approx \lambda_R v_R^2 - \mu_{\delta_R}^2, &\quad\quad\quad M_{\delta_L}^2 &\approx \lambda_{LR} v_R^2 -\mu_{\delta_L}^2
\end{array}
\label{LRmasses}
\end{align}
with the approximation that $M_{W_R} \gg M_{W_L}$ and $M_{Z_R} \gg M_{Z_L}$ and their mixing is negligible~\cite{Brehmer:2015cia}.

\subsection{Gauge coupling unification}

Much of the motivation for Pati-Salam models comes from their ultraviolet completion in $SO(10)$ or $E_6$, where all fermions are unified into a single representation of the group~\cite{Fritzsch:1974nn,Georgi:1974my,Gursey:1975ki}. Although we will not worry about the details of the UV completion beyond the GUT scale, we enforce the unification of the gauge couplings as it implies a relation between the different energy scales which is the source of complementarity between low energy and gravitational wave searches.

The one-loop gauge Renormalization Group Equations (RGEs) for the gauge couplings are
\begin{equation}
    \mu\frac{d g_a}{d\mu} = \frac{b_a}{16\pi^2}g_a^3.
    \label{gaugerges}
\end{equation}
for $a$ each element in a direct product of Lie groups, and $b_a$ a parameter that controls the slope of the RGE flow for $g_a$ and depends on group properties and the field content as~\cite{Machacek:1983tz}
\begin{equation}
  b_a = \frac{2}{3} \sum_{f}S(\mathcal{R}^a_f)d_\bot(\mathcal{R}^a_f) + \frac{1}{3} \sum_s S(\mathcal{R}^a_s) d_\bot(\mathcal{R}^a_s) - \frac{11}{3} C_2(\mathcal{G}_a)
\end{equation}
where $C_2(\mathcal{G}_a)$ is the Casimir of the group $\mathcal{G}_a$, $\mathcal{R}_f$ and $\mathcal{R}_s$ the representations of fermion and spinor fields, respectively, $S(\mathcal{R}_i)$ is the Dynkin index of the representation $\mathcal{R}_i$ and $d_\bot(\mathcal{R}_i)$ its dimension in the groups orthogonal to $\mathcal{G}_a$.

Equation~\eqref{gaugerges} can be solved analytically for each step of the breaking chain, and iterated from $M_{GUT}$ to $M_Z$ by using matching conditions at each scale. In fact for $\alpha_a= \frac{g_a^2}{4\pi}$ and $t = \frac{1}{2\pi} \log\mu$ the solution becomes a linear system of equations of the form
\begin{equation}
    \alpha_i^{-1}(M_Z) = \alpha_{GUT}^{-1} + \sum_{j=1}^m b_i^j \Delta t_j
\end{equation}
for $m$ steps in the breaking chain, $i = 1,2,3$ labels the SM gauge couplings and $\Delta t_j = t_j - t_{j-1}$.

The mechanism to achieve gauge coupling unification described above relies on the assumption that after each symmetry breaking, there is an Effective Field Theory remaining where many of the fields of the full theory have been integrated out, and these fields have all masses equal to the symmetry breaking scale. This is not generally the case and thus the matching conditions at each energy scale depend on the masses of these fields through the threshold corrections~\cite{Hall:1980kf, Weinberg:1980wa}
\begin{equation}
    \alpha^{-1}_i(\mu) = \alpha^{-1}_j(\mu) - \lambda_{ij}(\mu)
\end{equation}
with $\alpha_j$ and $\alpha_i$ the couplings of the theory before and after SSB, respectively. The threshold corrections at each scale can be computed as~\cite{Schwichtenberg:2018cka}
\begin{align}
    \lambda_{ij}(\mu) = \frac{1}{12\pi} \Bigg( \,& (C_2(\mathcal{G}_j) - C_2(\mathcal{G}_i)) - 21 \sum_g S(\mathcal{R}_g) \log \frac{M_g}{\mu}  \notag \\
    &+ 8 \sum_f S(\mathcal{R}_f) \log \frac{M_f}{\mu} +  \sum_s S(\mathcal{R}_s) \log \frac{M_s}{\mu}\Bigg)
\end{align}
where $g$, $f$ and $s$ label the vector, fermion and scalar fields integrated out at $\mu$, and $M_i$ are their masses.

The masses of the fields in the intermediate scales, eqs.\eqref{PSmasses}-\eqref{LRmasses}, are mostly fixed by the symmetry breaking conditions and the structure of the potential, and these contribute towards the threshold corrections whenever their masses stray from the respective scales at which they are integrated out. The mass of the field $\Omega_R$, integrated out at the PS scale, causes considerable threshold corrections since its mass is dominated by its portal coupling to $\Xi_1$, as it will be seen later when we discuss gravitational waves. Lastly most the masses of the fields at the GUT scale are unconstrained, since they depend on the field content and SSB mechanism at the GUT scale, which we do not consider here. Hence we take the liberty of setting these masses to values that assist in achieve gauge coupling unification within the desired ranges of relevant mass scales.

After adding threshold corrections we find that in our scenario with gauge coupling unification can be achieved for any value of $M_{PS} \in (2.9\times 10^3, 2.25 \times 10^7)$ GeV, and the remaining scales and couplings can be obtained in terms of $M_{PS}$. The ranges for other relevant quantities are
\begin{align}
    M_{LR}(M_{PS}) &\in (90.2, ~2.25 \times 10^7)\text{ GeV} \notag \\
    M_{GUT}(M_{PS}) &\in (1.4 \times  10^{16},~2.9 \times 10^{16}) \text{ GeV} \notag \\
    g_4(M_{PS})  &\in (0.75, 1.01) 
    \label{ranges}
\end{align}

\begin{figure}[t]
\centering
\includegraphics[width=\textwidth]{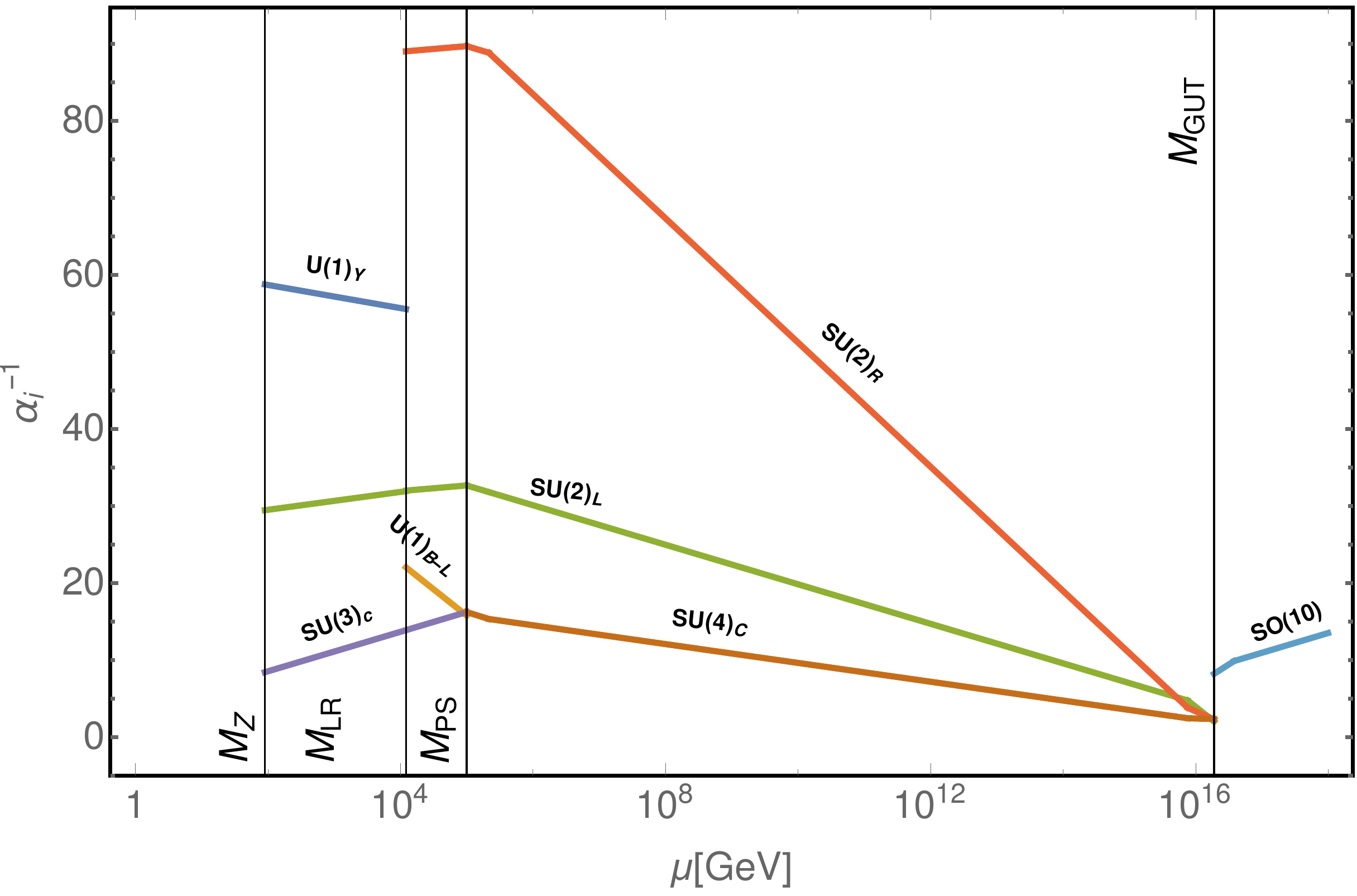}
\caption{Gauge coupling running for the model for $M_{PS} = 5\times 10^5$ GeV at one loop. The changes in slopes around $M_{PS}$ and $M_{GUT}$ and discontinuities at $M_{GUT}$ are due to threshold corrections.}
\label{fig:PSRGEs}
\end{figure}

Fig.~\ref{fig:PSRGEs} shows the one-loop RGE evolution of the gauge couplings in this model, for the choice of $M_{PS} = 10^5$ GeV, which we will later motivate as the optimal choice for the detection of gravitational waves. As can be noticed in the figure, around $M_{PS}$ and $M_{GUT}$ there are changes in slopes and discontinuities in the matching on the gauge couplings. These are a consequence of the threshold corrections described above, where the strong effect of the $SO(10)$ corrections can be readily spotted. 

\subsection{Thermal potential}
As described in the previous subsections, the Pati-Salam symmetry is broken when $\Xi _{1}$ acquires a vacuum expectation value. 
In the absense of new fermions charged under $SU(4)_C$, the resulting phase transition can be described by the Pati-Salam gauge coupling, the portal couplings between $\Xi _{1}$ and other scalars, and the scalar field potential in the $\Xi _{1}$ direction. 
The only portal coupling that can be large without undesirable low energy consequences is the mixed quartic between $\Xi _{1}$ and $\Omega _R $.
Here we will consider this term to set the effective mass of the  $\Omega_R$ field. As described in the previous subsection, the $SU(4)_C$ gauge coupling constant is fixed as a function of the PS mass. Then, our parameter space is limited to the two parameters in the $\Xi_1$ potential, and a single portal coupling.\footnote{We consider only renormalizable operators, and leave the case of Pati-Salam phase transitions in the presence of large non-renormalizable operators in the scalar potential to future work.}
With these considerations in mind we can approximate the scalar potential as follows
  \begin{align}
     V_{\Xi_1} &= -\mu_{\Xi_1}^2 \Xi_1^\dagger \Xi_1 + \lambda_1 [\Xi_1^\dagger \Xi_1]^2+ \Xi_1^\dagger \Xi_1 \left( \rho_1\Omega_R \Omega_R^\dagger \right) .
      \end{align}
It is convenient to reparametrize the zero-temperature scalar potential in terms of the overall scale and vev,
\begin{equation}
    V_0 = \Lambda ^4 \left( -\frac{1}{2} \left( \frac{\phi}{v} \right)^2 + \frac{1}{4} \left( \frac{\phi}{v} \right)^4   + \frac{\rho_1}{2} \frac{v^2}{\Lambda^4} \left( \frac{\phi}{v} \right)^2 \Omega_R \Omega_R^\dagger \right) \ ,
\label{eq:zerotemp}
\end{equation}
with $\mu_{\Xi_1}^2 = \lambda_1 v^2$, $\lambda_1 = (\Lambda/v)^4$ and $\Xi_1^v = \phi$.
We will see that the strength of the phase transition can be effectively determined based on the zero-temperature ratio $v/\Lambda$ and the size of the portal coupling $\rho_1$ \cite{Croon:2018erz}.
  
At one loop, we consider Coleman-Weinberg contributions and thermal corrections to the potential \eqref{eq:zerotemp},
\begin{eqnarray}
    V_{T \neq 0} &=&  \sum _{i} \frac{T^4}{2 \pi ^2} n_i J_B\left( \frac{m_i^2+\Pi _i }{T^2}  \right) \\
    V_{\rm CW} &=&n_{\rm GB} \frac{m_{\rm GB}^4}{64 \pi ^2}  \left( \log \left[ \frac{m_{\rm GB}^2}{\mu^2} \right] -\frac{5}{6} \right) + \sum _{i \neq \text{GB}} n_i \frac{m_i ^4}{64 \pi^2} \left( \log \left[ \frac{m_i^2}{\mu^2} \right] -\frac{3}{2} \right) \nonumber .
\end{eqnarray}
In the above equation, the sum is over all bosons ($n_i$ denotes the multiplicity factors), $\rm GB$ refers to gauge bosons, $\mu$ is the renormalization scale (in our analysis, we will assume $\mu \sim T$)\footnote{Alternatively, one could have chosen $\mu \sim M_{PS}$. This choice gives numerically and qualitatively similar results.}, and the mass terms are field dependent and given by\footnote{Radiative corrections to the tree level masses due to CW contributions have no appreciable effect at the target accuracy.} 
\begin{eqnarray}
 m_\phi^2 &=& 3 \frac{\Lambda ^4}{v^4} \phi ^2 - \frac{\Lambda ^4}{v^2} \\
 m_{\rm G}^2 &=&\frac{\Lambda ^4}{v^4} \phi ^2 - \frac{\Lambda ^4}{v^2} \\ 
 m_{\rm GB}^2 &=& \frac{g_4^2}{6 } \phi ^2 \\
 m_{\Omega _R}^2  &\approx & \rho_1 \phi ^2 \ ,
\end{eqnarray}
for the physical field $\phi$, Goldstone modes $\rm G$, gauge bosons $\rm GB$ and the scalar $\Omega _R$ respectively.
We also include Debye masses, given by \cite{Parwani:1991gq}, to delay the breakdown of perturbation theory at high temperature. 
The Debye masses can be approximated in the high temperature limit as\footnote{Going beyond the high temperature limit for the Debye masses requires solving a self consistency condition, which is outlined in ref \cite{Curtin:2016urg}.} 
\begin{eqnarray}
\Pi _\phi &=&  \frac{3}{4} \frac{\Lambda ^4}{v^4} T^2 +\frac{g^2_4}{8} T^2 + \frac{45}{12} \rho _1 T^2\\ 
\Pi _{\rm GB} &=& \frac{47 g^2_4}{36} T^2 .
\end{eqnarray}
Note that we have assumed the combination of the Debye mass and the mass parameter, $\Pi_{\Omega _R}-\mu^2$, is negligible compared to the field dependent mass $m_{\Omega _R}$.
In the next sections, we use the full 1-loop thermal potential in the $\phi$ direction, 
\begin{equation}\label{eq:fullpot}
    V (\phi,T)= V_0 (\phi) + V_{\rm CW} (\phi,\mu) + V_{T \neq 0} (\phi,T)
\end{equation} to find the thermal parameters of the phase transition. 

\begin{figure}[t]
    \centering
    \includegraphics[width=.45\textwidth]{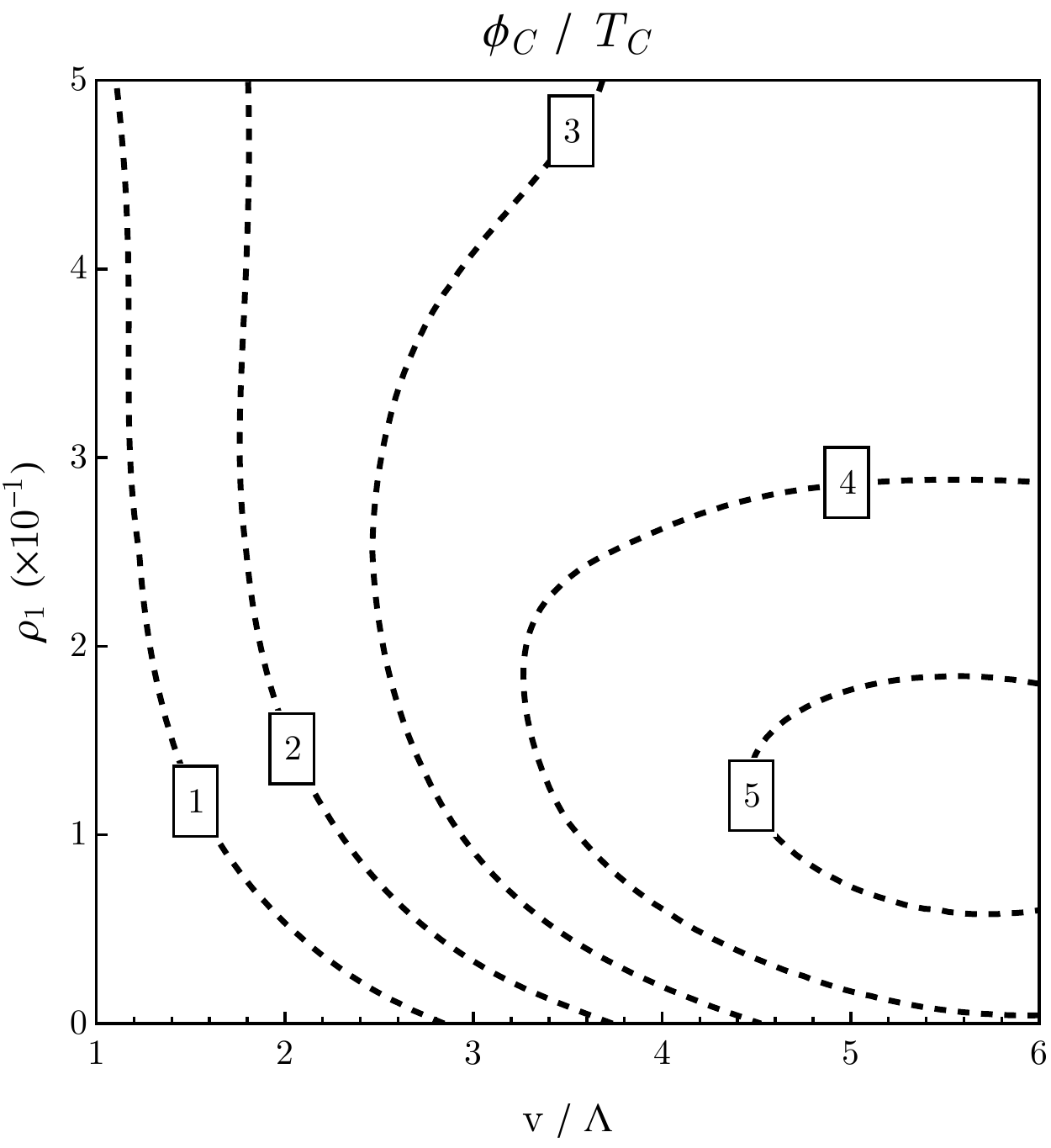}
    \caption{Strength of the phase transition expressed in the parameters $\phi_C/T_C$, as a function of the tree-level combination $v/\Lambda$ and the portal coupling $\rho_1$. In this plot, $M_{PS} = 10^5 $ GeV. For simplicitly, all other portal couplings have been set to zero.}
    \label{fig:strength}
\end{figure}

\section{Gravitational Wave spectrum from a PS phase transition}
\label{sec:PT}

\subsection{Strength of the phase transition}
At high temperature $T \gg v$, the scalar potential \eqref{eq:fullpot} has a single minimum at $\Xi=0$. As the sector cools, a second minimum develops with $\Xi \neq 0$. The minima are degenerate at the critical temperature, \begin{equation}
    V_{T_C}(0) = V_{T_C}(\phi_C).
\end{equation} 
Some intuition for the strength of the gravitational wave signal can be developed from the ratio $\phi_C/T_C$ for different parameter choices. In Fig.~\ref{fig:strength} this ratio is shown for different values of the portal coupling $\rho_1$ the ratio of zero temperature variables $v/\Lambda$. Here we have fixed the gauge coupling according to the relation in the previous subsection \eqref{ranges} and set $M_{PS} = 10^5$ GeV. It is seen that the ratio peaks at around $\phi_C/T_C = 5$ for large $v/\Lambda$, and portal coupling strength of around $\rho_1 \sim 10^{-1}$. This can be understood in the following way: as $\rho_1$ increases, two competing effects occur. The thermal mass term increases, such that the critical temperature is lower. At the same time, the Coleman-Weinberg potential contributes an effective interaction term, which drives the value of $\phi_C$ smaller. The CW potential depends on $\rho_1^2$, while the thermal potential depends on $\rho_1$ in the high temperature limit. Ultimately, the behavior of $\phi_C/T_C$ results from the balance of the two effects.

\subsection{Gravitational Wave spectrum}
To find the thermal parameters governing the phase transition, we find classical solutions to the Euclidean equations of motion (the scalar bounce solution), which describe the nucleation of an $O(3)$ bubble of the true vacuum in a medium of the false vacuum. We solve the Euclidean equations of motion by varying the initial conditions via a simple bisection method. We give more details on our calculation of the thermal parameters in appendix \ref{sec:Thermalparameters}. The thermal parameters can be used to predict the stochastic gravitational wave spectrum using a combination of analytic and lattice studies, as reviewed in appendix \ref{sec:GWspectra}.

\begin{figure}[ht]
    \centering
    \includegraphics[width=0.32\textwidth]{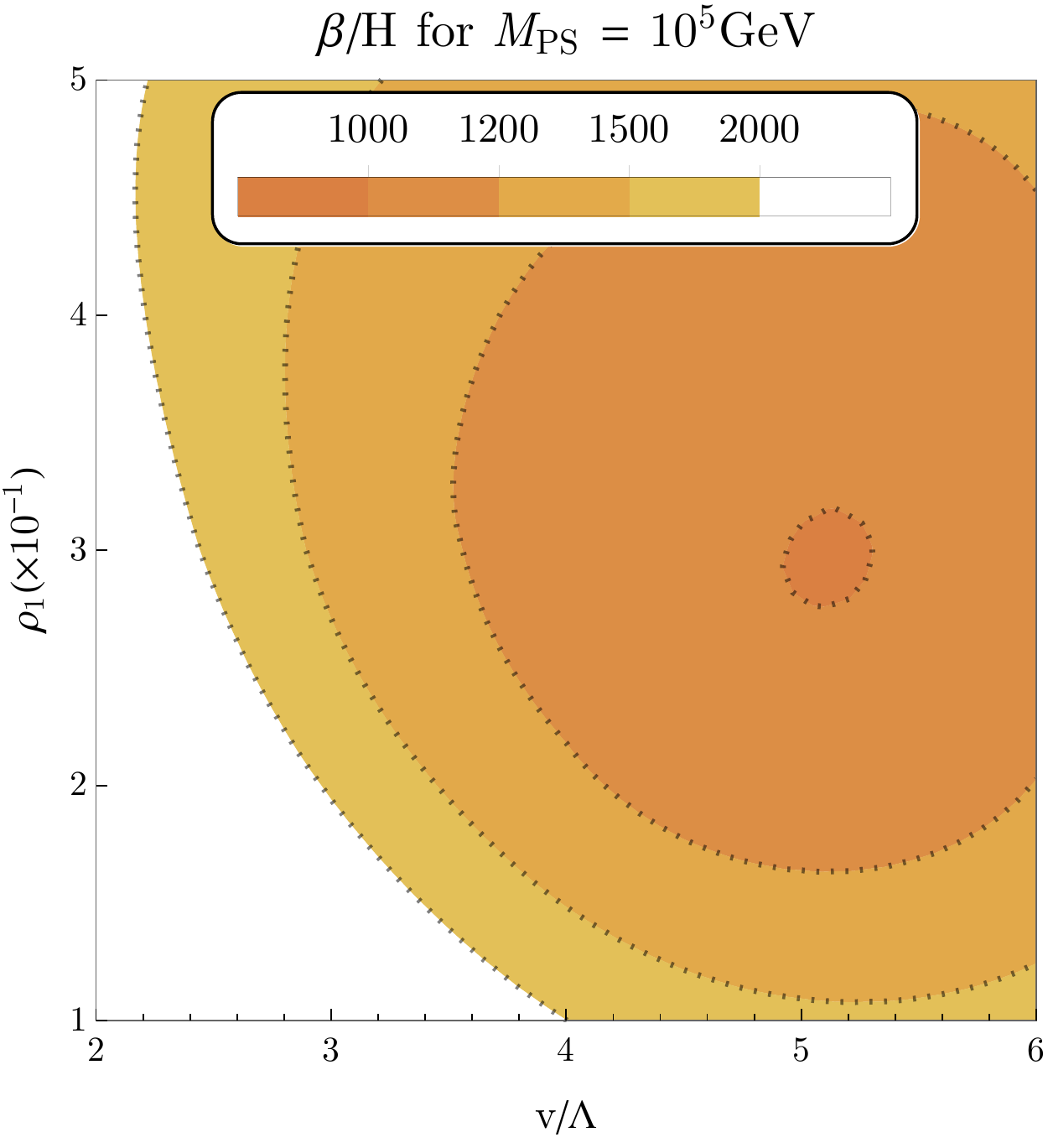}
    \includegraphics[width=0.32\textwidth]{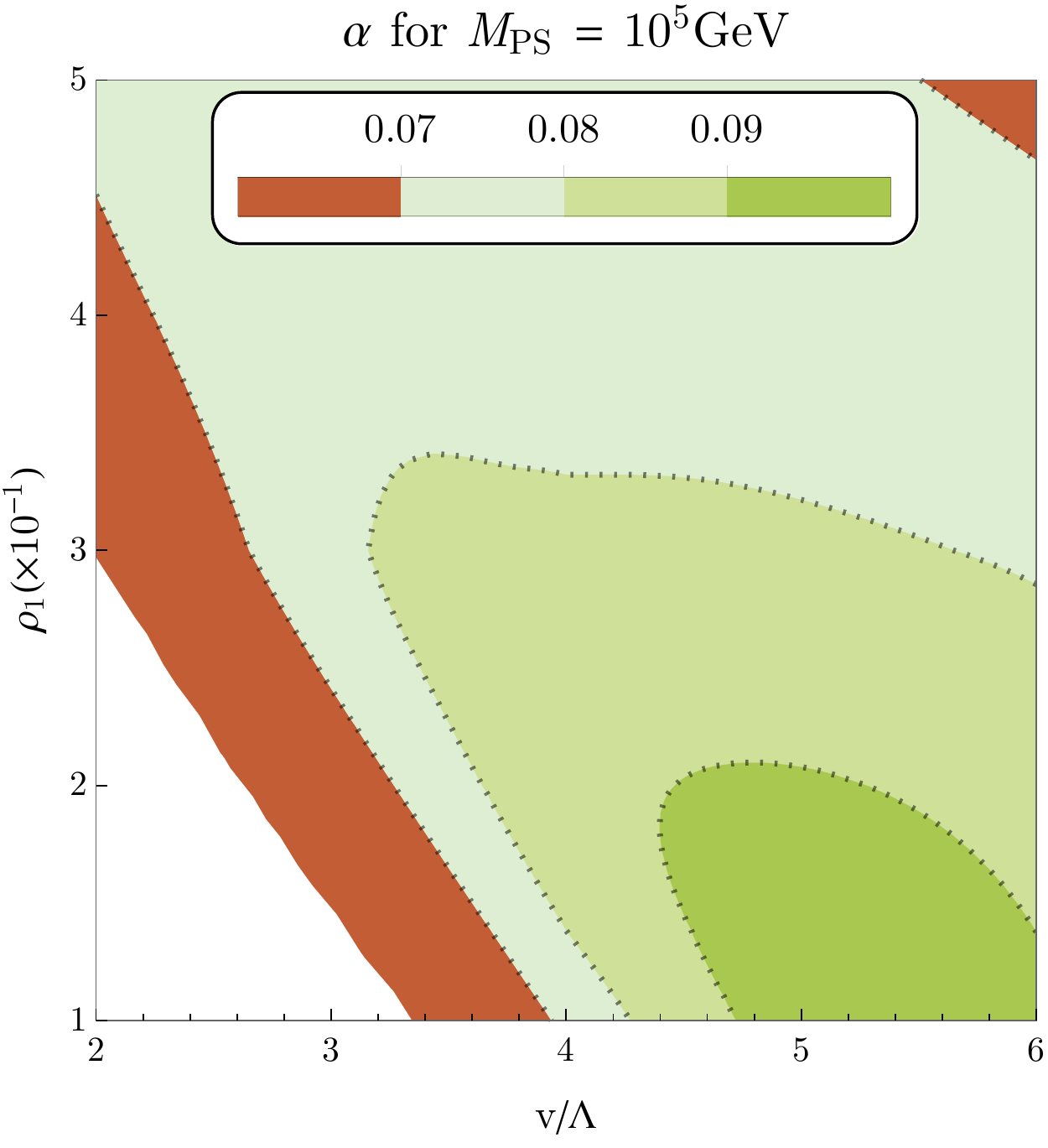}
    \includegraphics[width=0.32\textwidth]{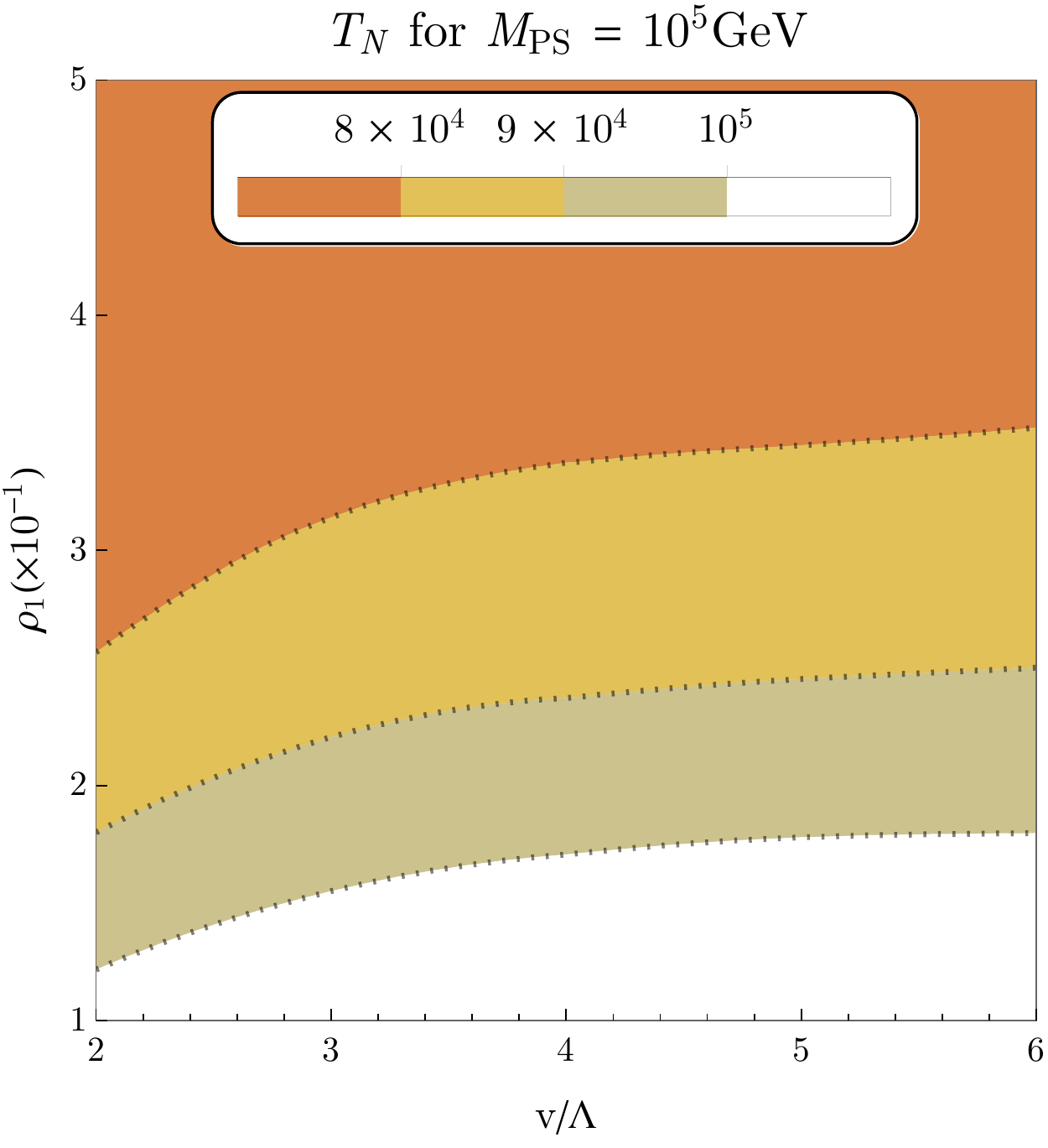}
    \caption{Behaviour of thermal parameters with Lagrangian parameters $\rho_1$ and $v/\Lambda$. The left panel shows the speed of the transition as captured by $\beta/H$. It is seen that $\beta/H$ is minimized for portal coupling $\sim 3 \times 10^{-1}$ and large $v/\Lambda \sim 5$. In the center, the latent heat which is largest for portal coupling $\sim 10^{-1}$ and large $v/\Lambda$. The right panel shows the nucleation temperature, which does not vary much as a result of the fixed scale $M_{PS} = 10^5$ GeV.}
    \label{fig:ampplot}
\end{figure}


Informed by the results in Fig.~\ref{fig:strength}, we vary the portal coupling between $0.1<\rho _1 < 0.5$ and the ratio of scales $2<v/\Lambda < 6$ and calculate the nucleation temperature $T_N$ as well as the thermal parameters $\alpha$ and $\beta/H$. The results of our parameter scan are smoothed using a local quadratic regression with tri-cube weights \cite{cleveland1988locally}.
We show the behaviour of the thermal parameters with $v/\Lambda$ and $\rho_1$ in Fig. \ref{fig:ampplot}.  It is seen that the latent heat $\alpha$ asymptotes for large $v/\Lambda$ and reaches a maximum for $\rho_1 \sim O(10^{-1})$ which is expected from the behaviour of the order parameter shown in Fig. \ref{fig:strength}. Likewise, the speed of the transition $\beta/H$ also reaches a minimum value for $v/\Lambda \sim 5$ and coupling $\rho_1 \sim 3 \times 10^{-1}$. The nucleation temperature is mostly determined by the scale $M_{PS}$, but shows a small dependence on the coupling strength $\rho_1$.

\begin{figure}
    \centering
    \includegraphics[width=0.7\textwidth]{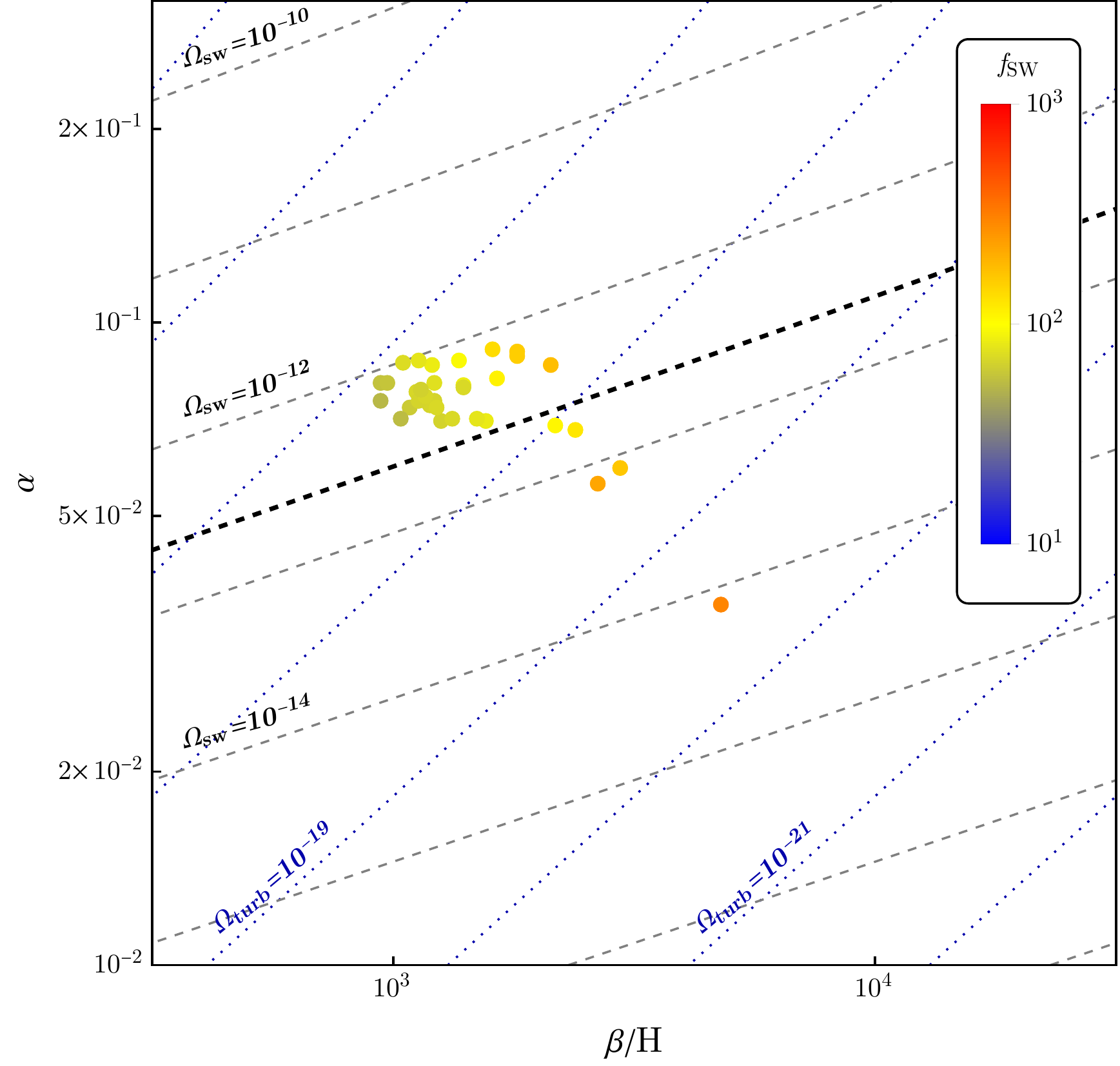}
    \caption{The sound wave (in black) and turbulence (in blue) spectra of a Pati-Salam phase transition for the benchmark points described in the text. Here we have assumed $v_w = 1$ as motivated in appendix \ref{sec:GWspectra}. To find the peak of the turbulence spectrum, which depends on two scales, we used the fiducial value $T_N = 10^5$ GeV. The colour scaling denotes the peak frequency for the sound wave spectrum; the peak frequency of the turbulence spectrum is expected to be smaller but of the same order of magnitude for these benchmarks. The thicker line shows the peak sensitivity of the Einstein telescope~\cite{Punturo:2010zz}.}
    \label{fig:benchmarks}
\end{figure}

The thermal parameters can be used to find the stochastic gravitational wave spectra, a calculation we review in appendix \ref{sec:GWspectra}. We plot contours of peak amplitude of the gravitational wave spectra in the $(\beta/H,\alpha)$ plane in Fig.~\ref{fig:benchmarks}, with a selection of benchmark points from our study. The benchmarks shown here represent an evenly spaced grid with $\rho_1 = \{1,2,3,4,5\} \times 10^{-1}$ and $v/\Lambda = \{2,3,4,5,6\}$. The colour scaling in this plot gives the frequency at the peak of the sound wave spectrum, which is the dominant contribution. 

For reference, the thicker dashed line in Fig.~\ref{fig:benchmarks} gives the anticipated peak sensitivity of the Einstein telescope \cite{Punturo:2010zz}. We expect the sound wave peak to be visible at the Einstein telescope \cite{Punturo:2010zz} for $\beta/H \sim O(10^3)$, $T_n \lesssim 10^5$ GeV and $\alpha \gtrsim 0.07$  \cite{Figueroa:2018xtu} with the Cosmic Explorer \cite{Evans:2016mbw} allowing slightly higher values of $(T_n \times \beta/H)$. 

\section{Complementarity with low energy probes}\label{sec:lowenergy}

A Pati-Salam model with $M_{PS} = 10^5$ GeV, such as studied in the previous section, has many other observational consequences. Through gauge coupling unification, a fixed value of $M_{PS}$ also fixes the remaining scales and the gauge couplings. The values of these for $M_{PS} = 10^5$ GeV are
\begin{align}
    M_{LR} &\sim 1.2 \times 10^4 \text{ GeV}, \notag \\
    M_{GUT} &\sim 1.91 \times 10^{16} \text{ GeV}, \notag \\
    g_4 &\sim 0.88, \notag \\
    g_{GUT} &\sim 0.96.
\end{align}

Using these values in this section we will study some low-energy signatures of this model, such as neutrino masses, lepton flavour violation, collider searches and proton decay.

\subsection{Neutrino masses}

As required by the observation of neutrino oscillations~\cite{Pontecorvo:1967fh, Fukuda:1998mi, Ahmad:2002jz, Eguchi:2002dm, Ahmad:2001an}, left-handed neutrinos have non-vanishing masses. However, studies of the CMB by the Planck satellite have imposed a strong upper limit on the sum of the neutrino masses $\sum m_\nu < 0.23$ eV~\cite{Ade:2015xua}. Left-right symmetric models, such as the intermediate step of the model described in Section~\ref{sec:model}, naturally contain a right-handed neutrino field $N$ and a left-handed triplet $\delta_L$, which can make the active neutrinos light via type I and type II seesaw mechanisms~\cite{Mohapatra:1979ia, Schechter:1980gr}. The mass matrix of neutrinos in this scenario is
\begin{equation}
    M_\nu = \left(\begin{matrix}M_L & M_D \\ M_D^T &M_R\end{matrix}\right),
\end{equation}
where $M_D$ is the Dirac-type mass of the neutrinos and $M_L$ and $M_R$ are the left and right-handed Majorana masses. The former mass arises from the Yukawa coupling of the neutrino field to the SM Higgs after electroweak symmetry breaking, whereas the latter masses are generated dynamically through the vacuum expectation values of $\delta_{L/R}$, denoted by $v_L$ and $v_R$ respectively. In the limit of small active-sterile mixing, we can write the light left-handed neutrino masses as
\begin{equation}
  m_{\nu_L} \simeq M_L - M_D M_R^{-1} M_D^T,
  \label{seesawIandII}
\end{equation}
and the masses of the heavy right-handed neutrinos is $m_{N} \simeq M_R$. This relation can be expressed in terms of the various vevs from Eq.~\eqref{vevs} by taking $M_L = \zeta v_L$, $M_R = \zeta v_R$ and $M_D = y_\nu v_{SM}$ as
\begin{equation}
    m_{\nu_L} \simeq \zeta v_L - y_\nu^2 \frac{v_{SM}^2}{\zeta v_R},
    \label{numasses}
\end{equation}
where $y_\nu$ is the Yukawa coupling of the neutrinos and $\zeta$ is the coupling of the scalar triplets $\delta_L$ and $\delta_R$ to the lepton fields, which we have taken to be equal, $\zeta_L = \zeta_R = \zeta$ as a relic of $D$ parity from the GUT scale. In models with breaking of LR manifest symmetry the vev $v_L$ can be obtained as a function of the scale of the scale of $D$-parity breaking~\cite{Chang:1983fu, Deppisch:2014zta, Deppisch:2015cua}. In this scenario, $D$-parity is broken already at the GUT scale, and the only coupling of $\delta_L$ to the field responsible is through $\Omega_R$. Hence we can express $v_L$ as~\cite{Deppisch:2014zta}
\begin{equation}\label{eq:vL}
    v_L \approx \frac{\eta}{M^2} \frac{v_{SM}^2 v_R M_{\Omega_R}^2}{M_{GUT}^2} \approx \frac{\eta \rho_1}{M^2} \frac{v_{SM}^2 v_R v^2}{M_{GUT}^2},
\end{equation}
with $\eta$ the coupling between $\delta_L$ and $\Omega_R$ (see Appendix \ref{sec:scalarpotential}) and $M$ the dimensionful coupling between $\Omega_R$ and the $D$-parity breaking field. Identifying $M_{LR} \equiv v_R$, we can rewrite eq.\eqref{numasses} as
\begin{equation}
    m_{\nu_L} \simeq \frac{v_{SM}^2}{M_{LR}} \left(\zeta \frac{\eta \rho_1}{M^2} \frac{ v^2 M_{LR}^2}{M_{GUT}^2} - \frac{y_\nu}{\zeta}\right).
    \label{numasses2}
\end{equation}

Electroweak precision data restricts $v_L \lesssim 5$ GeV in order to keep the electroweak $\rho$ parameter under control~\cite{Kanemura:2012rs}. This translates into an upper limit on $\frac{\eta}{M^2} \lesssim 8.2 \times 10^{14}\text{ GeV}^{-1}$. The parameter $M$ controls the splitting between the GUT scale and the mass of $\Omega_R$, which we want to remain at around $M_{PS}$ (c.f. eq.~\eqref{PSmasses}). $M$ is then constrained as $M \lesssim 1.6\times10^{-12}$ GeV and it thus forces a strong upper bound on $\eta \lesssim 2.03 \times 10^{-9}$.

The parameters $\zeta$ and $y_\nu$ are unconstrained in this model, but $M_{LR}$ depends on the Pati-Salam scale $M_{PS}$ via gauge coupling unification. For the gravitational wave scenario studied in the previous section, $M_{PS} = 10^5$ GeV, $M_{LR} \sim 12$ TeV and $\rho_1 \sim 0.3$. For the maximum values allowed for $M$ and $\eta$, the condition that the sum of neutrino masses is below the CMB limit~\cite{Ade:2015xua} becomes
\begin{equation}
    y_\nu > \zeta (-1.54 \times 10^{-11} + 1.001 \zeta)
\end{equation}

There is another contribution to neutrino masses that we have not considered here arising from loop corrections involving heavy leptoquarks~\cite{Dorsner:2017wwn}. These contributions are rather small and do not modify the conclusions of Eq.~\eqref{numasses2} significantly. Therefore we will not discuss them any further.

The mixing of sterile to active neutrinos $\Theta$ is the source of many of the contributions from heavy neutrinos to low energy observables, including electroweak precision observables (EWPO), which set a upper limit of $|\Theta|^2 \lesssim 10^{-3}$. This mixing is given by $\Theta = M_D M_R^{-1}$, c.f. eq.~\eqref{seesawIandII}, which using the parameters $\zeta$ and $y_\nu$ transforms into
\begin{equation}
    y_\nu < 1.562 ~\zeta.
\end{equation}

\subsection{Lepton flavour violation}

Neutral lepton flavour violation is present in the Standard Model, through oscillations of neutrinos via their mixing matrix~\cite{Pontecorvo:1959sn,Maki:1962mu}. Charged lepton flavour violation, however, cannot be mediated in the SM, and so any observation of these processes would be a smoking gun for BSM physics~\cite{Kuno:1999jp, Gluza:2016qqv, Deppisch:2012vj,FileviezPerez:2017zwm}.

Many decay and conversion processes have been studied that violate lepton flavour, such as the photonic penguins, $\mu \to e\gamma$, $\tau \to e \gamma$, $\tau \to \mu \gamma$, three-body penguins and box diagrams, $l^- \to l^- l^+ l^-$ and $\mu -e$ conversion in nuclei~\cite{Abada:2014kba}. Searches for these processes have been performed by several experiments and they have set upper limits on their branching ratios ~\cite{TheMEG:2016wtm,Aubert:2009ag,Hayasaka:2007vc,Bellgardt:1987du,Lees:2010ez,Hayasaka:2010np,Aad:2016wce,Aaij:2014azz,Kaulard:1998rb,Honecker:1996zf,Bertl:2006up}. The most constraining of these are the limits on $\mu \to e \gamma$, by the MEG collaboration~\cite{TheMEG:2016wtm}, $\mu \to eee$ , by the SINDRUM experiment~\cite{Bellgardt:1987du}, and $\mu - e$ conversion in nucleii, by SINDRUM-II ~\cite{Bertl:2006up}, which are
\begin{align}
\notag BR(\mu \to e \gamma) &< 4.2 \times 10^{-13}, \\
\notag BR(\mu \to eee) &< 1.0 \times 10^{-12}, \\
R^{Au}(\mu - e) &< 8 \times 10^{-13}.
\label{lfvexp}
\end{align}

In left-right symmetric models the gauge bosons $W_R$ and the scalars $\delta_{L,R}$ can mediate these processes, and one can approximate their branching fractions as~\cite{Cirigliano:2004mv, Deppisch:2014zta}
\begin{align}
    BR(\mu \to e\gamma) &\sim 1.5\times 10^{-7} |\Theta_{e I}^*\Theta_{\mu I}|^2 \left(\frac{g_R}{g_L}\right)^4 \left(\frac{m_{N}}{m_{W_R}}\right)^4  \left(\frac{1\text{ TeV}}{M_{W_R}}\right)^4, \notag \\
    BR(\mu \to eee) &\sim \frac{1}{2} |\Theta_{e I}^*\Theta_{\mu I}|^2|\Theta_{e I}|^4 \left(\frac{g_R}{g_L}\right)^4 \left(\frac{m_{N}}{m_{W_R}}\right)^4  \left(\frac{M_{W_R}^4}{M_{\delta_R}^4} + \frac{M_{W_R}^4}{M_{\delta_L}^4}\right), \notag \\
    R^N(\mu -e) &\sim0.73\times 10^{-9}  X_N |\Theta_{e I}^*\Theta_{\mu I}|^2
    \left(\frac{g_R}{g_L}\right)^4 \left(\frac{m_{N}}{m_{W_R}}\right)^4  \left(\frac{1\text{ TeV}}{M_{\delta_R}}\right)^4 \left(\log\frac{m_{\delta_R}^2}{m_\mu^2}\right)^2.
\end{align}
where the nuclear form factor $X_N$ has the value $X_{Au} = 1.6$~\cite{Deppisch:2014zta} and $\Theta$ is the active-sterile neutrino mixing matrix.

In this model we have made the simplifying assumption that $M_{\delta_L} \sim M_{\delta_R} \sim M_{LR}$, and also we have that $M_{W_R} = \tfrac{1}{2} g_R M_{LR}$ and $M_N = \zeta M_{LR}$. If we take the scenario that maximizes the detection of gravitational waves, $M_{PS} = 10^5$ GeV, then $M_{LR} = 12.15 \times 10^3$ GeV, $g_L = 0.627$ and $g_R = 0.376$. The active-sterile mixings are not fixed by the scenario, but electroweak precision data has put an upper limit on their values which, as a conservative limit, we can take as $|\Theta_{\alpha I}|^2  < 10^{-3}$~\cite{Drewes:2015iva}. This results in the branching ratios
\begin{align}
BR(\mu \to e\gamma) &\sim 3.43\times10^{-10} \zeta^4, \notag \\
BR(\mu \to eee) &\sim 6.15\times10^{-9} \zeta^4, \notag \\
R^N(\mu -e) &\sim 2.71\times10^{-12} \zeta^4,
\end{align}
which then gives an upper limit for $\zeta$ so as to satisfy the limits in eq.~\ref{lfvexp}, $\zeta <0.113$, due to the most constraining of the observables, namely $\mu \to eee$. 

%
%

Future experiments measuring $\mu -e$ conversion, such as COMET~\cite{Kurup:2011zza} and Mu2e~\cite{Kutschke:2011ux} aim to reach the limit of $R^N(\mu -e) < 10^{-16}$. A positive signal from either of those experiments would fix $\zeta$ for this model, which would strengthen our predictions and motivation for the complementarity with gravitational wave detection. If no such signal is found, the new limits would further constrain the value of $\zeta$. Taking the form factor $X_{Al} = 0.8$~\cite{Deppisch:2014zta} this new upper limit would drop to $\zeta < 0.093$.

\subsection{Collider Searches}

Low scale Pati-Salam and left-right symmetric models predict light exotic particles that can be visible at the LHC. In particular, the lightest exotic states produced in our model after LR symmetry breaking are the heavy right-handed neutrinos $N_j$, the right-handed gauge bosons $W_R$ and $Z_R$ and the left-handed scalar triplet $\delta_L$, with masses in \eqref{LRmasses}.

Right handed neutrinos can be produced directly on-shell at the LHC from the decay of a $W_L$ boson. The primary process for detection of right-handed neutrinos at the LHC is $pp \to W \to Nl \to Wll \to lljj$, where the two final state leptons have the same sign~\cite{Deppisch:2015qwa}. ATLAS and CMS reported exclusion limits on searches for same sign dilepton final states for mass ranges of 100 GeV $< M_N < 500$ GeV ~\cite{Aad:2015xaa} and 20 GeV $< M_N < 1600$ GeV~\cite{Sirunyan:2018xiv}, respectively. In the heavy mass range, above the $Z$ resonance, $M_N > 90$ GeV, CMS has the strongest exclusion power which is almost linear in the $M_N - |\Theta_{eN}|^2$ plane, so its limit can be approximated as
\begin{equation}
    \frac{M_N}{|\Theta_{eN}|^2} \gtrsim 1.5 \text{ TeV}.
\end{equation}

This can be translated to the parameters of our model using the relations $M_N \eqsim \zeta v_R$ and $\Theta_{eN} \sim \frac{y_\nu v_{SM}}{\zeta v_R}$ and for the chosen value of $M_{PS} = 10^5$ GeV, as
\begin{equation}
    y_\nu^2 < 1.98 \times 10^{4} \zeta^3 \quad \text{for } \zeta > 7.4 \times 10^{-3}.
\end{equation}

In addition to the same sign dilepton search, CMS reported results on searches for heavy neutrinos in three lepton final states~\cite{Sirunyan:2018mtv}. The limits of the search for $M_N > 100$ GeV are rather similar to the dilepton search. For smaller masses below the $Z$ resonance, the exclusion limits of this search are among the strongest in the literature on par with the results of DELPHI~\cite{Abreu:1996pa}, and it effectively excludes all neutrino masses for $\Theta_{eN} > 10^{-5}$. So we have the constraint
\begin{equation}
    y_\nu < 0.156~ \zeta \quad \text{for } \zeta < 7.4 \times 10^{-3}.
\end{equation}

In the case that the gauge boson $W_R$ can be produced at the LHC, another channel opens for the production of right-handed neutrinos where the $W_R$ takes the place of $W_L$ in the decay chain. In this channel, the two leptons in the final states can have either the same or opposite signs, depending on the Majorana or Dirac nature of the neutrinos~\cite{Deppisch:2015cua}. Both ATLAS and CMS reported strong exclusion limits for $W_R$ and $N$ in searches with two same and opposite sign leptons and two jets final states~\cite{Aaboud:2018spl,Sirunyan:2018pom}. The limits from both experiments reach up to $M_{W_R} > 4.7$ TeV for 500 GeV$ < M_N < 3$ TeV, for the simplified model where $g_L = g_R$ and maximal coupling $|\Theta| = 1$.

However, in the cases where $g_L \neq g_R$, as it is our model, the constraint is slightly relaxed. In order to assess the effect of these searches on our model we make a very rough comparison of the number of events predicted in our model for the same-sign $eejj$ signal region with the measured data by ALTAS and CMS at $36~\rm{fb}^{-1}$~\cite{Aaboud:2018spl,Sirunyan:2018pom}. The cross-section of this model for this process can be estimated to be (to leading order in $\zeta$)\footnote{We use the expressions mentioned in \cite{Deppisch:2015cua} for the production cross section of $W_R$ as well as the branching ratios of $W_R$ and $N$.}
\begin{equation}
    \sigma(pp \to W_R \to eejj) \approx  5.356 ~ y_\nu^{-2} ~\zeta^4~ \rm{fb}
\end{equation}

ATLAS and CMS reported a number of observed events of 11 and 4, and predicted background events 11.2 and 2.6, respectively. Using the reported efficiencies for the high $W_R$ mass region of 0.54 (ATLAS) and 0.57 (CMS) we find that, at 95\% CL 
\begin{equation}
\begin{array}{lll}
    y_\nu &< 4.010~ \zeta^2 -77.222~ \zeta^4 &(\rm{ATLAS}) , \\
    y_\nu &< 4.500~ \zeta^2 - 86.539~ \zeta^4 &(\rm{CMS}).
\end{array}
\end{equation}

The other heavy gauge boson in the theory with a mass low enough to be relevant for collider searches is $Z_R$. High mass resonance searches in the dilepton invariant mass from ATLAS and CMS have imposed strong constraints on the mass of $Z_R$~ \cite{Aaboud:2017buh, CMS:2016abv}. Both experiments give a simplified model limit in the range $M_{Z_R} > (3.5, 4.0) $ TeV. It has been shown that for models with $g_L \neq g_R$ the limits on $Z'$ resonances are much weaker~\cite{Patra:2015bga}. In any case, for $M_{PS} = 10^5$ GeV, the mass of $M_{Z_R}$ in our model is fixed to $M_{Z_R} \approx 14.5$ TeV and hence it is not affected from the current experimental limits.

Finally let us consider searches for triplet scalar bosons $\delta_L$. Most relevant to us are searches for doubly charged scalar bosons, $\delta_L^{++}$, in same sign diboson final states~\cite{Chiang:2012dk}. The lower limits set by such searches are of the order of a few hundred GeV, depending on the vev $v_L$. Similarly to the $Z_R$ case above, for $M_{PS} = 10^5$ GeV, $M_{\delta_L} \approx M_{LR} = 12.152$ TeV, and thus the limits do not affect the outcome of this model.

So far the LHC experiments have reported analyses on $36 ~\rm{fb}^{-1}$ of collected data. The increased sensitivity of future upgrades of the LHC will impose stronger constraints on the masses of exotic states. The high-luminosity LHC (HL-LHC) is projected to collect up to $1~ \rm{ab}^{-1}$ of data at 14 TeV, and the hypothetical upgrade to the Very Large Hadron Collider (VLHC) will push the energy frontier to 100 TeV with a projected luminosity of $10 ~\rm{ab}^{-1}$~\cite{Ruiz:2017nip}. Discovery of any of $N$, $W_R$, $Z_R$ or $\delta_L$ in either HL-LHC or VLHC would conclude in strong evidence towards a LR symmetric model at low scales, which motivates a low scale PS phase transition leading to a GW spectra observable in the next generation of experiments.

\subsection{Proton decay}

Unified theories typically introduce baryon number violating operators, which render the proton unstable and can lead to rapid proton decay~\cite{Marciano:1981un, Abbott:1980zj, Langacker:1980js}. This is certainly true for $SO(10)$ models, whose off-diagonal gauge and scalar bosons couple to both quarks and leptons an can mediate nucleon decays~\cite{Dutta:2004zh,Kolesova:2014mfa, Deppisch:2017xhv}. This is not the case, however, for Pati-Salam models, where the gauge sector preserves baryon and lepton number independently and only selected scalar sectors can mediate the transition, none of which we include in our model~\cite{Mohapatra:1980qe}.

Therefore, the only source of proton decay in our model arises from the leptoquarks at the GUT scale. The half-life of the proton in this scenario, with mass $m_p$, then can be approximated as~\cite{Tanabashi:2018oca, Deppisch:2017xhv}
\begin{equation}
    \tau_p \approx \frac{(4\pi)^2}{\lambda_X^4}\frac{M_{X}^4}{m_p^5},
    \label{protondecay}
\end{equation}
where $M_X \sim M_{GUT}$ is the mass scale of the mediator and $\lambda_X$ its coupling to the quarks and leptons, which corresponds to $g_{GUT}$ for a gauge mediator.

Proton decay transitions can occur in a number of different channels, e.g. $p \to e^+\pi^0$, $p \to e^+ K^0$, etc.~\cite{Nath:2006ut,Abe:2014mwa}. The most constraining limit was imposed by the Super-Kamiokande collaboration to the process $p \to e^+ \pi^0$, with a half-life lower bound of $\tau_p > 1.29 \times 10^{34}$ years~\cite{Nishino:2012bnw}.

In our particular scenario, with the optimal PS scale for gravitational wave detection, $M_{PS} = 10^5$ GeV, gauge coupling unification fixes $M_{GUT} \sim 1.9\times 10^{16}$ and $g_{GUT} \sim 0.96$, which gives a proton half-life of $\tau_p \sim 7\times10^{35}$ years, larger than the experimental limit. This prediction for proton decay is not too far from the SuperK bound and in fact it is fairly close to the projected limit expected to be reached by HyperK~\cite{Abe:2015zbg} of $\tau > 1.3 \times 10^{35}$ years. A positive measurement of proton decay is the smoking gun for unified theories, in particular if the measured decay rate is close to the predicted in our model, it would further motivate the scenario with $M_{PS} = 10^5$ GeV where the peak amplitude of GW spectra is within sensitivity of the Einstein telescope. Otherwise, if proton decay is not observed, a stronger upper limit of the half-life of the proton would fall within range of our prediction and therefore a more detailed calculation of the decay rate and RGE evolution would need to be performed in order to assess the survivability of the model.

\begin{figure}[ht]
    \centering
    \includegraphics[width=0.7\textwidth]{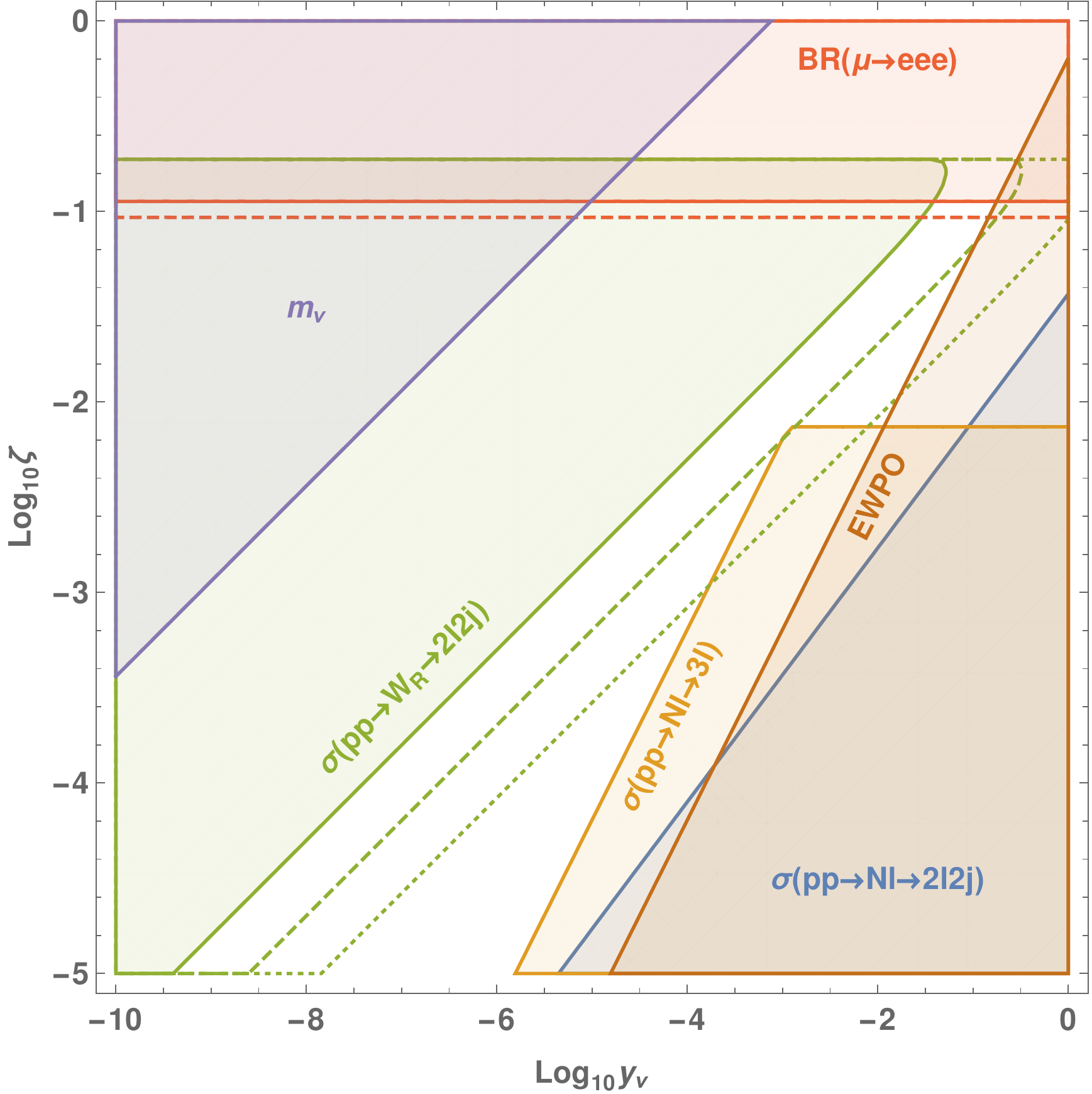}
    \caption{Exclusion limits on the parameters $\zeta$ and $y_\nu$ by the searches for heavy neutrinos in $lljj$ (blue) and $3l$ final states (orange), searches for $W_R$ bosons (green), LFV (red), EWPO (brown) and the cosmological limit on neutrino masses (purple). Dashed and dotted lines mark the expected sensitivity of future experiments. Here we have set $M_{PS} = 10^5$ GeV which in turn determines $M_{LR} \sim 1.2 \times 10^4$ GeV.}
    \label{fig:lowenergy}
\end{figure}

\section{Discussion and conclusion}
\label{sec:discussion}

The Pati-Salam phase transition is a unique candidate for a gravitational wave spectrum which peaks within the frequency windows of ground-based interferometer experiments. If such a signal is observed, complementarity with low-energy experiments can be used to probe the Pati-Salam parameter space.

The strength of the phase transition and the corresponding gravitational wave signal depend most importantly on the degrees of freedom with a large coupling to the broken direction. 
Therefore, we considered an effective model with four free parameters: the PS gauge coupling $g_4$, the PS scale $M_{PS}$, the portal coupling $\rho_1$, and the ratio of parameters in the tree-level scalar potential $v/\Lambda$. We found that an observation of a broken power-law spectrum of gravitational waves which peaks for $f \sim [10-1000]$ can be explained by a Pati-Salam model with scale $M_{PS} \sim 10^5$ GeV. An argument from gauge coupling unification fixes the Pati-Salam coupling $g_4$ as a function of this scale. The amplitude of the power spectrum, then, is a function of the portal coupling and the zero-temperature combination ($v/\Lambda$). As was demonstrated in Section~\ref{sec:PT}, the peak of the spectrum may be observable if at least one of the portal couplings is sizable ($\rho_1 \gtrsim 0.1$), and for particular zero-temperature parameters in the scalar potential ($v/\Lambda \gtrsim 2$).

For a Pati-Salam scale of $M_{PS}\sim 10^5$ GeV, many of the low energy constraints described in Section~\ref{sec:lowenergy} impose limits on the parameters $y_\nu$ and $\zeta$.\footnote{In addition to these limits, low-energy neutrino constraints have a weak dependence on the scalar portal coupling $\rho_1$ through Eq.~\eqref{eq:vL}.} We summarize these constraints in Fig.~\ref{fig:lowenergy}, including collider constraints for decays of heavy neutrinos (blue and orange), decays of $W_R$ (green), the cosmological limit on the neutrino masses (purple), LFV constraints (red) and the limit from EWPO (brown).\footnote{The calculation of many of these constraints was done in using approximate methods, so these exclusion limits are subject to a more precise analysis which we leave to the subject of future investigation.} As can be seen in the figure, a large part of the parameter space is excluded by several searches. However, there is still a narrow band where this model parameters are allowed. The future projections of several experiments are depicted with dashed and dotted lines, with the projected limit on $\mu - e$ conversion from COMET and Mu2e in dashed red, and the limits for $W_R$ searches in dashed green (HL-LHC) and dotted green (VLHC). These future searches will be able to explore the parameter space more thoroughly and further constrain the model. The included set of low energy probes is but a subset of the possible relevant phenomenological observables of PS and LR models, which we have chosen to elucidate the complementarity with GW searches. We defer the computation of other relevant observables, such has neutrinoless double beta decay or electric dipole moments to further work.

Finally we note that phase transitions in models of Grand Unified Theories are often associated with the formation of cosmic defects. One-dimensional defects, cosmic strings, may decay into gravitational radiation \cite{Kibble:1976sj,Kibble:1980mv}. However, cosmic strings associated with the energy scales studied in this work - $M_{PS} \sim 10^5$ GeV - will have a dimensionless string tension of $G\mu \sim 10^{-29}$ and will therefore not lead to any observational signatures. Primordial monopoles created during the PS phase transition can be reduced to acceptable limits during late time inflation~\cite{Okada:2013vxa}. Light monopoles may be produced at colliders~\cite{Kephart:2017esj}, however PS monopoles have a mass of the order $\sim 10^{6}$ GeV~\cite{Lazarides:1980va} which is beyond the reach of current experimental searches by ATLAS~\cite{Aad:2015kta} and MoEDAL~\cite{Acharya:2016ukt}.

\section*{Acknowledgements}
The authors would like to thank Q. Shafi, Y. Zhang, and D. Weir for useful discussions.
TEG was partly funded by the Research Council of Norway under FRIPRO project number 230546/F20 and partly supported by the ARC Centre of Excellence for Particle Physics at the Tera-scale, grant CE110001004. TRIUMF receives federal funding via a contribution agreement with the National Research Council of Canada and the Natural Science and Engineering Research Council of Canada. 

\appendix

\section{Scalar potential}
\label{sec:scalarpotential}

The scalar potential of the Pati-Salam model at zero temperature can be written as
\begin{align}
    V_0 &= V_{\Xi_1} + V_{\Xi_2} + V_{\Xi_2} + V_{\Omega_R} + V_{\Delta_L} + V_{\Delta_R} + V_{\Phi} \notag \\
        &+ V_{\Xi\Xi} + V_{\Xi\Psi} + V_{\Omega_R\Psi} + V_{\Delta\Delta} + V_{\Delta\Phi}
\end{align}
where $V_\Psi$ refers to the terms in the potential that contain only the field $\Psi$, and these are
\begin{equation}
    V_{\Psi} = -\mu_{\Psi}^2 \mathrm{Tr}[\Psi^\dagger\Psi] + \lambda_{\Psi} |\mathrm{Tr}[\Psi^\dagger\Psi]|^2 + \lambda_{\Psi}' \mathrm{Tr}[\Psi^\dagger \Psi\Psi^\dagger \Psi],
\end{equation}
with $\Psi = \Xi_1, \Xi_2, \Xi_3, \Omega_R, \Delta_L, \Delta_R$. Since the fundamental representation of $SU(2)$ is real, the field $\tilde{\Phi} = \tau_2 \Phi^* \tau_2$ transforms as $\Phi$. Henceforth we call $\Phi_1 = \Phi$ and $\Phi_2 = \tilde{\Phi}$. The self-interaction term $V_\Phi$ now looks like
\begin{equation}
    V_{\Phi} = \sum_{ij}-\mu_{ij}^2 \mathrm{Tr}[\Phi_i^\dagger\Phi_j] + \sum_{ijkl}\lambda_{ijkl} \mathrm{Tr}[\Phi_i^\dagger\Phi_j]\mathrm{Tr}[\Phi_k^\dagger\Phi_l] + \lambda_{ijkl}' \mathrm{Tr}[\Phi_i^\dagger \Phi_j\Phi_k^\dagger \Phi_l].
\end{equation}
The term $V_{\Xi\Xi}$ contains interactions among $\Xi_{(1,2,3)}$ of the form
\begin{equation}
    V_{\Xi\Xi} = \sum_{ijkl} \lambda_{ijkl}\mathrm{Tr}[\Xi_i^\dagger\Xi_j]\mathrm{Tr}[\Xi_k^\dagger\Xi_l] +  \lambda_{ijkl}'\mathrm{Tr}[\Xi_i^\dagger\Xi_j\Xi_k^\dagger\Xi_l]
\end{equation}
where $(i,j,k,l) = (1,2,3)$ and not all $i,j,k,l$ are equal. The term $V_{\Xi\Psi}$ contains the portal couplings of the fields $\Xi_i$ with the rest and they are of the type
\begin{align}
    V_{\Xi\Psi} = \sum_i &~\mathrm{Tr}[\Xi_i^\dagger \Xi_i] \left(\rho_{i1}\mathrm{Tr}[\Omega_R^\dagger \Omega_R] + \rho_{i2}\mathrm{Tr}[\Delta_L^\dagger\Delta_L] + \rho_{i3}\mathrm{Tr}[\Delta_R^\dagger \Delta_R] + \sum_{jk}\rho_{ijk}\mathrm{Tr}[\Phi_j^\dagger\Phi_k]\right) \notag \\
    &+ \mathrm{Tr}[\Xi_i^\dagger \Xi_i \left(\rho_{i1}'\Omega_R^\dagger \Omega_R + \rho_{i2}'\Delta_L^\dagger\Delta_L + \rho_{i3}'\Delta_R^\dagger \Delta_R + \sum_{jk}\rho_{ijk}'\Phi_j^\dagger\Phi_k\right)] \notag \\
    &+ \rho_{i}''\mathrm{Tr}[\Xi_i^\dagger\Omega_R\Delta_R^\dagger \Delta_R] + \sum_{jk}\rho_{ijk}''\mathrm{Tr}[\Xi_i^\dagger \Omega_R \Phi_j^\dagger \Phi_k].
\end{align}
The term $V_{\Omega\Psi}$ is fairly similar to $V_{\Xi\Psi}$ and looks like
\begin{align}
    V_{\Omega\Psi} &=\mathrm{Tr}[\Omega_R^\dagger \Omega_R] \left( \eta_{1}\mathrm{Tr}[\Delta_L^\dagger\Delta_L] + \eta_{2}\mathrm{Tr}[\Delta_R^\dagger \Delta_R] + \sum_{jk}\eta_{ijk}\mathrm{Tr}[\Phi_j^\dagger\Phi_k]\right) \notag \\
    &+ \mathrm{Tr}[\Omega_R^\dagger \Omega_R \left(\eta_{1}'\Delta_L^\dagger\Delta_L + \eta_{2}'\Delta_R^\dagger \Delta_R + \sum_{jk}\eta_{ijk}'\Phi_j^\dagger\Phi_k\right)].
\end{align}
The last two terms, $V_{\Delta\Delta}$ and $V_{\Delta\Phi}$ have the same form as in LR symmetric models
\begin{align}
    V_{\Delta\Delta} &= \lambda_{\Delta} \mathrm{Tr}[\Delta_L^\dagger\Delta_L] \mathrm{Tr}[\Delta_R^\dagger \Delta_R]  + \lambda_{\Delta}' \mathrm{Tr}[\Delta_L^\dagger \Delta_L \Delta_R^\dagger \Delta_R], \\
     V_{\Delta\Phi} &= \sum_{ij}\mathrm{Tr}[\Phi_i^\dagger \Phi_j] ~\left(\lambda_{Lij} \mathrm{Tr}[\Delta_L^\dagger \Delta_L] + \lambda_{Rij} \mathrm{Tr}[\Delta_R^\dagger\Delta_R]\right) \notag \\
     &+ \sum_{ij}\mathrm{Tr}[\Phi_i^\dagger\Phi_j\left(\lambda_{Lij}'\Delta_L^\dagger\Delta_L + \lambda_{Rij}'\Delta_R^\dagger\Delta_R + \lambda_{LRij} \Delta_L^\dagger \Delta_R\right)].
\end{align}

This scalar potential contains all possible terms allowed by the gauge symmetries. In the work above we have chosen to remove a few of them setting their couplings to zero, e.g. all portal couplings vanish $\rho_{ij} = \rho_{ij}' = \rho_{ij}'' = \rho_{ijk} = \rho_{ijk}'= 0$ save for the first one, that we have renamed in the text as $\rho_{11} = \rho_1 \neq 0$.

\section{Thermal parameters} \label{sec:Thermalparameters}

The nucleation temperature, which approximates the collision temperature very well when the phase transition occurs quickly, is conventionally defined as the temperature for which a volume fraction $e^{-1}$ is in the true vacuum state. This corresponds approximately to
\begin{equation}
    p(t_N) t_N^4 = 1
\end{equation}  
where $p(t)$ is the nucleation probability per unit time per unit volume, and where $t_N$ is the nucleation time. The nucleation probability can be calculated from the bounce solution as,
\begin{equation}
    p(T) = T^4 \,e^{-S_E/T}
\end{equation} 
where $S_E$ is the Euclidean action evaluated on the bounce which approximates a tanh function and can be solved by bisection or perturbing a tanh ansatz \cite{Akula:2016gpl,White:2016nbo}.\footnote{In our analysis, we assume a radiation dominated universe to relate the nucleation temperature and time. See, however, \cite{Ellis:2018mja}.}

The speed of the phase transition can be calculated from the rate of change of the euclidean action
\begin{equation}
    \frac{\beta}{H} = T \frac{d (S_E/T)}{dT}
\end{equation}
Lastly, the most important parameter governing the amplitude of the relic gravitational waves will be the latent heat released in the transition (normalized to the radiation density)
\begin{equation}
    \alpha = \left. \frac{\Delta V- T \Delta dV/dT}{\rho _*}  \right|_{T_n}
\end{equation}
where $\rho _* = \pi^2 g_* T^4/30$. 

\section{Gravitational Wave Spectrum}\label{sec:GWspectra}
The gravitational wave spectrum from a cosmic phase transition can be expressed as a sum of three contributions,
\begin{equation}
    \Omega _{GW}(f)  h^2 = \Omega _{\rm coll}(f)h^2 + \Omega _{\rm sw}(f)h^2 +\Omega _{\rm turb} (f) h^2
\end{equation}
denoting contribution from collision of scalar shells, the collision of the sound shells and the turbulence respectively. 
Lattice simulations indicate that all three spectra can be captured by a broken power law, with a peak frequency and amplitude dependent on the thermal parameters at collision: $(T_*, \beta/H , v_w)$ and $(\beta/H, \alpha , v_w)$ respectively. 
The collision term is expected to dominate for so-called runaway bubble walls whose Lorentz boost factor approaches infinity. 
It was recently realized that vacuum transitions in which gauge bosons gain a mass are not expected to runaway \cite{Bodeker:2017cim}. This is confirmed by simple condition that the mean field potential lifts the PS breaking minimum above the symmetric one \cite{Bodeker:2009qy}.
However, $v_w$ at collision is still expected to be large, and in our analysis we use $v_w \rightarrow 1$.

For non-runaway transitions, the sound wave contribution is expected to dominate \cite{Hindmarsh:2016lnk,Hindmarsh:2017gnf}, although recent work has suggested that lattice simulations may overestimate this contribution \cite{Ellis:2018mja}. In this work, we calculate the thermal parameters from first principles and consider the peak sound wave amplitude analytically fitted to lattice simulations to be an approximation of the GW spectrum. It is given by \cite{Weir:2017wfa},
\begin{equation}
h^2\Omega _{\rm sw}    = 8.5 \times 10^{-6} \left( \frac{100}{g_*} \right)^{-1/3} \Gamma ^2 \bar{U}_f^4  \left( \frac{\beta}{H} \right)^{-1}  v_w S_{\rm col}(f) 
\end{equation}
where $\bar{U}_f^2\sim (3/4) \kappa _f \alpha$ is the rms fluid velocity and $\Gamma \sim 4/3$ is the adiabatic index. For $v_w \rightarrow 1$, the efficiency parameter is well approximated by \cite{Espinosa:2010hh},
\begin{equation}
    \kappa _f \sim \frac{\alpha }{0.73+0.083 \sqrt{\alpha } +\alpha}
\end{equation}
and the spectral shape is
\begin{equation}
    S_{\rm sw} =  \left( \frac{f}{f_{\rm sw}} \right) ^3 \left( \frac{7}{4+3\left( \frac{f}{f_{\rm sw}}\right) ^2} \right)^{7/2}
\end{equation}
with peak frequency
\begin{equation}
    f_{\rm sw} = 8.9 \times 10^{-7} {\rm Hz} \left( \frac{z_p}{10} \right) \frac{1}{v_w} \left( \frac{\beta}{H} \right) \left( \frac{T_N}{{\rm Gev}} \right) \left( \frac{g_* }{100} \right)^{1/6} \ ,
\end{equation}
where $z_p$ is a simulation derived factor which we take to be $6.9$ from \cite{Weir:2017wfa}. The power spectrum from the turbulence contribution is
\begin{eqnarray}
 h^2 \Omega _{\rm turb} = 3.354 \times 10^{-4} \left( \frac{\beta}{H} \right)^{-1} \left( \frac{\kappa \epsilon \alpha }{(1+\alpha } \right)^{3/2} \left( \frac{100}{g^*} \right) ^{1/3}v_w S_{\rm turb} (f) 
\end{eqnarray}
where $\epsilon$ is the fraction of the energy in the plasma is expressed as turbulence; in our results, we use $\epsilon = 0.05$. The spectral form is given by
\begin{equation}
    S_{turb} = \frac{(f/f_{\rm turb})^3}{[1+(f/f_{\rm turb})]^{11/3}(1+\frac{8 \pi f}{h_*})} \ . 
\end{equation}
The Hubble rate at the transition temperature as well as the peak frequency are given by
\begin{eqnarray}
 h_* &=& 16.5 \mu {\rm Hz} \left( \frac{T_N}{100 {\rm GeV}} \right) \left( \frac{g^*}{100} \right)^{1/6}  \\
 f_{\rm turb } &=& 27 \mu {\rm Hz} \frac{1}{v_w}  \left( \frac{T_N}{100 {\rm GeV}} \right) \frac{\beta }{H}  \left( \frac{g^*}{100} \right)^{1/6}
\end{eqnarray}
respectively.
Recent work \cite{Ellis:2018mja} has shown that it is difficult to satisfy the criteria that the phase transition completes and the sound waves last longer than a Hubble time. The consequence of this is that the sound waves are likely overestimated and the turbulence is likely underestimated. The suppression of the sound wave peak is naively estimated to be suppressed by a factor \cite{Ellis:2018mja}
\begin{equation}
\frac{H \bar{R}}{\bar{U}_f} \sim
    \left( \frac{\beta}{H} \right)^{-1} \times \alpha^{-1} \times \frac{(8 \pi )^{1/3}}{\frac{3}{4} \kappa _f } = \left[6 - 7.5 \right]
\end{equation}
where we have used the relation for the rms fluid velocity $U_F \sim \frac{3}{4} \kappa _f \alpha $. The last equality holds for the points in our scan, which have $0.07<\alpha <0.09$. However, the precise suppression factor is subject to future lattice simulations. Similarly, \cite{Ellis:2018mja} argued that the turbulence factor has been underestimated, however much uncertainty remains about the precise form of the turbulence spectrum in general.

\bibliography{references}
\end{document}